\documentclass[prb,reprint,aps,showpacs,
unsortedaddress,
superscriptaddress,floatfix,longbibliography]{revtex4-1}
\pdfoutput=1

\usepackage[english]{babel}
\usepackage[utf8]{inputenc}
\usepackage{newtxtext, newtxmath}
\usepackage[scaled=0.8]{beramono}
\usepackage[T1]{fontenc}
\usepackage{breqn}
\usepackage{dutchcal} 
\usepackage[normalem]{ulem}

\usepackage{acronym}
\usepackage{bm}
\usepackage{dsfont}
\usepackage{graphicx}
\usepackage{mathtools}
\usepackage{microtype}
\usepackage{tikz}

\usetikzlibrary{decorations.markings}
\usetikzlibrary{decorations.pathmorphing}

\definecolor{mauve}{RGB}{94, 60, 153}

\usepackage[colorlinks, allcolors=mauve]{hyperref}
\usepackage[figure, figure*]{hypcap}

\let\Pi\varPi

\tikzset{%
    baseline=-0.43ex,
    el/.style={
        line cap=round,
        decoration={markings, mark=at position 1/2 with {\arrow[xshift=1.5pt]>}},
        preaction={decorate},
        thick,
        },
    el_gray/.style={
        line cap=round,
        decoration={markings, mark=at position 1/2 with {\arrow[xshift=1.5pt]>}},
        preaction={decorate},
        ultra thick, color=gray
        },
    ph/.style={
        line cap=rect,
        decoration={snake, segment length=3mm, pre length=1mm, post length=1mm},
        decorate,
        thick,
        },
    elel/.style={
        line cap=rect,
        decoration={snake, aspect=0, pre length=0.3mm, post length=0.3mm},
        decorate,
        thick,
        },
    elhbubble/.pic={
        \fill[white] (0, 0) to[out=-45, in=-135] (2, 0) to[out=135, in=45] cycle;
        \draw[el, ultra thick] (0, 0) to[out=-45, in=-135] (2, 0);
        \draw[el, ultra thick] (2, 0) to[out=+135, in=+45] (0, 0);
        },
    elhbubble_thin/.pic={
        \fill[white] (0, 0) to[out=-45, in=-135] (2, 0) to[out=135, in=45] cycle;
        \draw[el] (0, 0) to[out=-45, in=-135] (2, 0);
        \draw[el] (2, 0) to[out=+135, in=+45] (0, 0);
        },
    elhbubble_anh/.pic={
        \fill[white] (0, 0) to[out=-45, in=-135] (2, 0) to[out=135, in=45] cycle;
        \draw[el] (0, 0) to[out=-45, in=-135] (2, 0);
        \draw[el] (2, 0) to[out=+135, in=+45] (0, 0);
        },
    elhbubble_/.pic={
        \fill[white] (0, 0) to[out=-45, in=-135] (2, 0) to[out=135, in=45] cycle;
        \draw[el, ultra thick, color = gray] (0, 0) to[out=-45, in=-135] (2, 0);
        \draw[el, ultra thick, color = gray] (2, 0) to[out=+135, in=+45] (0, 0);
        },
    bubble/.pic={
        \draw[ph,double] (0, 0) to[out=120, in=80] (2, 0);
        },
    bubble_bare/.pic={
        \draw[ph] (0, 0) to[out=120, in=80] (2, 0);
        \draw[el](0,0) to (2,0);
        },
    chi/.pic={\pic{bubble}; \node at (1, 0.25) {$\Sigma (T, \omega)$};
        },
    chiPH/.pic={\pic{elhbubble}; \node at (1, 0.05) {$\Pi (T, \omega)$};
        },
    chiPH_anh/.pic={\pic{elhbubble_anh}; \node at (1, 0.05) {$\pi^{anh} (T, \omega=0)$};
        },
    chi_bare/.pic={\pic{bubble_bare}; \node at (1, 0.25) {$\Sigma (T, \omega)$};
        },
    chiPH_bare/.pic={\pic{elhbubble_thin}; \node at (1, 0.05) {$\Pi (T, \omega)$};
        },
    chi0/.pic={\pic{bubble}; \node at (1, 0.05) {$\chi^b (\sigma, 0)$};
        },   
    chis/.pic={
        \draw[el] (0, 0) to[out=-70, in=-110] (1, 0);
        \draw[el] (1, 0) to[out=+110, in=+70] (0, 0);
        },
    Db/.pic={
        \draw[ph] (0, 0) -- (1, 0);
        \useasboundingbox (-0.15, 0) -- (1.15, 0);
        \node[below=1mm] at (0.5, 0) {$D^0$};
        },
    D/.pic={
        \draw[ph, double] (0, 0) -- (1, 0);
        \useasboundingbox (-0.15, 0) -- (1.15, 0);
        \node[below=1mm] at (0.5, 0) {$D(T, \omega)$};
        },
    D2/.pic={
        \draw[ph, double] (0, 0) -- (1, 0);
        \useasboundingbox (-0.15, 0) -- (1.15, 0);
        \node[below=1mm] at (0.5, 0) {$D^{\text{anh}}$};
        },
    Gb/.pic={
        \draw[el] (0, 0) -- (1, 0);
        \useasboundingbox (-0.15, 0) -- (1.15, 0);
        \node[below=1mm] at (0.5, 0) {$G^0$};
        },
    Gp/.pic={
        \draw[el, double] (0, 0) -- (1, 0);
        \useasboundingbox (-0.15, 0) -- (1.15, 0);
        \node[below=1mm] at (0.5, 0) {$G^p (T, \omega)$};
        },
    G/.pic={
        \draw[el, ultra thick] (0, 0) -- (1, 0);
        \useasboundingbox (-0.15, 0) -- (1.15, 0);
        \node[below=1mm] at (0.5, 0) {$G(T, \omega)$};
        },
    G_arrow/.pic={
        \draw[el, ultra thick] (0, 0) -- (2, 0);
        \useasboundingbox (-0.15, 0) -- (1.15, 0);
        },
    v/.pic={
        \draw[elel] (0, 0) -- (1, 0);
        \node[below=2pt] at (0.5, 0) {$v$};
        \useasboundingbox (-0.15, 0) -- (1.15, 0);
        },
    U/.pic={
        \draw[elel, double] (0, 0) -- (1, 0);
        \useasboundingbox (-0.15, 0) -- (1.15, 0);
        \node[below=2pt] at (0.5, 0) {$U(T, \omega)$};
        },
    U0/.pic={
        \draw[elel, double] (0, 0) -- (1, 0);
        \useasboundingbox (-0.15, 0) -- (1.15, 0);
        \node[below=2pt] at (0.5, 0) {$U(\sigma, 0)$};
        },
    W/.pic={
        \draw[elel, ultra thick] (0, 0) -- (1, 0);
        \useasboundingbox (-0.15, 0) -- (1.15, 0);
        \node[below=2pt] at (0.5, 0) {$W(T, \omega)$};
        },
    ob/.pic={
        \fill (0, 0) circle (2pt);
        },
    o/.pic={\fill (0, 0) circle (3pt);},
    op/.pic={
        \draw[thick, fill=white] (0, 0) circle (20pt); \node at (0, 0) {$\pi^{\text{anh}}$};
        },
    op2/.pic={
        \draw[thick, fill=white] (0, 0) circle (20pt); \node at (0, 0) ;
        },
    gb/.pic={
        \pic{ob}; 
        \node[above=2pt];
        },
    gr/.pic={
        \pic {o};
        \node[above=2pt] ;
        }}

\newacro{BZ}{Brillouin zone}
\newacro{cDFPT}{constrained density-functional perturbation theory}
\newacro{cRPA}{constrained random-phase approximation}
\newacro{HLRN}{North-German Supercomputing Alliance}
\newacro{DFG}{Deutsche Forschungsgemeinschaft}
\newacro{DFPT}{density-functional perturbation theory}
\newacro{DFT}{density-functional theory}
\newacro{LDA}{local-density approximation}
\newacro{NLCC}{nonlinear core correction}
\newacro{PBE}{Perdew-Burke-Ernzerhof}
\newacro{RPA}{random-phase approximation}
\newacro{SC}{semicore}
\newacro{EPC}{electron-phonon interaction}

\def\Zagreb{Centre for Advanced Laser Techniques, Institute of Physics, 10000 Zagreb, Croatia}

\def\IRB{Ru\dj er Bo\v{s}kovi\'c Institute, 10 000 Zagreb, Croatia}

\def\Modena{Dipartimento di Scienze Fisiche, Informatiche e Matematiche,
Universit\`a di Modena e Reggio Emilia, Via Campi 213/a I-41125 Modena, Italy}

\def\Modenaa{Centro S3, Istituto Nanoscienze-CNR, Via Campi 213/a, I-41125 Modena, Italy}

\def\Liege{Nanomat/Q-MAT/CESAM and European Theoretical Spectroscopy Facility, Universit\'e de Li\`ege, B-4000, Li\`ege, Belgium}

\def\Utrecht{ITP, Physics Department, Utrecht University 3508 TA Utrecht, The Netherlands}

\begin{document}

\sloppy

\title{Electron-mediated anharmonicity and its role in the Raman spectrum of graphene}

\author{Nina Girotto Erhardt}
\affiliation\Zagreb

\author{Alo\"is Castellano}
\affiliation\Liege

\author{J. P.
Alvarinhas Batista}
\affiliation\Liege

\author{Raffaello Bianco}
\affiliation\Modena
\affiliation\Modenaa

\author{Ivor Lon\v{c}ari\'c}
\affiliation\IRB

\author{Matthieu J. Verstraete}
\affiliation\Liege
\affiliation\Utrecht

\author{Dino Novko}
\email{dino.novko@gmail.com}
\affiliation\Zagreb

\begin{abstract}
The Raman active G mode in graphene exhibits a strong coupling to electrons, yet the comprehensive treatment of this interaction in the calculation of its temperature-dependent Raman spectrum remains incomplete. In this study, we calculate the temperature dependence of the G-mode frequency and linewidth, and successfully explain the experimental trend, by accounting for the contributions arising from the first-order electron-phonon coupling, electron-mediated phonon-phonon coupling, and standard lattice anharmonicity. The generality of our approach enables its broad applicability to study phonon dynamics in materials where both electron-phonon coupling and anharmonicity are important. 
\end{abstract}

\maketitle

\section*{Introduction}
Understanding phonons and their interactions is crucial to elucidate the role of lattice dynamics in material properties\,\cite{giustino17}, such as carrier and lattice transport, optical response, superconductivity, and structural phase transitions. 
Raman spectroscopy, a powerful and widely applicable experimental technique, plays a key role in studying temperature-dependent phonon behavior. This technique is particularly powerful for atomically thin materials like graphene and transition metal dichalcogenides~\cite{ferrari13,zhang2015}, where the position of the phonon peaks can reveal the actual doping concentration as well as the number of layers~\cite{das08,tan2011,zhao2013}. 
These experiments highlight the crucial role of electron-phonon coupling (EPC) and its competition with phonon-phonon interaction, showing how the interplay between phonons and electronic excitations  influences the Raman features of layered materials~\cite{Rafailov02, Yoon2013, sohier19, sarkar20, paul2021}. 
In this regard, the case of graphene is especially intriguing~\cite{Berciaud2010,ferrante18,howard11}.

Raman spectroscopy is also a very important probe for \emph{dynamical} EPC effects, which are expected to be crucial when the electron and phonon excitations are close in energy\,\cite{maksimov96,engelsberg63}. These non-adiabatic (NA) effects have been explored, both experimentally and theoretically, in transition metal dichalcogenides, diamond, magnesium diboride (MgB$_2$) and various single-layer materials~\cite{caruso17,cappelluti06,ponosov16,ponosov98,novko2018a,novko2020a,eiguren20,nina23,novko20b}, but graphene remains the benchmark example of the failure of the static Born-Oppenheimer approximation and strong dynamical EPC~\cite{pisana07,lazzeri06,saitta08,piscanec07,meng22a,meng22b}. The graphene G mode (an optical phonon with E$_{2g}$ symmetry), is its most prominent Raman feature. It corresponds to an in-plane vibration of carbon atoms and strongly couples to electrons and other phonons~\cite{bonini07}. Raman spectroscopy of graphene has been extensively studied~\cite{Linas15,calizo07,Shaina2016,Yoon11,TIAN2016,Lee2017,Ferrari06,tuinstra70,EVERALL1991,PingHeng99,sonntag23,liu19,hlinka07,torres12,Casiraghi09}, but the results are very sensitive to the experimental conditions and different groups report different linewidths and frequency behaviour as a function of temperature. However, all agree that the G mode frequency decreases with temperature in a monotonous manner. Experimental evidence for the linewidth range from its slow increase~\cite{lin11,Han22} to two different temperature-dependent regimes~\cite{chae10} resembling the Raman measurements performed on non-equilibrium and photo-doped graphene samples~\cite{ferrante18,Tiberj2013} and Weyl semimetals~\cite{Coulter19}.

One of the open questions in the Raman spectrum of graphene concerns the source of the observed temperature dependence for the G mode frequency and linewidth. 
From a theoretical point of view, four-phonon anharmonicity partially explains the observed G mode temperature trends in graphene and graphite~\cite{bonini07,Han22,lin11,MONTAGNAC13,chen21,giura12}. 
On the other hand, Raman experiments done under ultrafast laser excitation, prove that electrons must be involved in the relaxation process in graphene~\cite{ferrante18}. This prompts us to question whether the G mode frequency decrease and an anomalous linewidth temperature behaviour could be the result of a combined effect of anharmonicity and EPC. In spite of its omnipresence, EPC is considered to have a weak contribution to the G mode linewidth at medium to strong doping: long-wavelength interband electron transitions are then prohibited by Pauli's exclusion principle ~\cite{Tsuneya06,chae10}. Contrary to experimental observations\,\cite{howard11,chen23}, this would imply that highly-doped graphene should have a zero EPC linewidth contribution, in spite of a strong doping-induced increase in overall EPC. For a long time it has been considered that the spectral signature of EPC should be a temperature-induced linewidth \emph{decrease}~\cite{ponosov98, ponosov05, lazzeri06, bonini07}. However, this belief was often based on first-order calculations of EPC, which may not be adequate to capture the relevant physics in this case. A similar debate arose for MgB$_2$.

The similarities between MgB$_2$ and graphene begin in their structure. Carbon atoms, arranged in a hexagonal pattern in graphene are effectively substituted with boron atoms in MgB$_2$. Meanwhile, magnesium atoms are positioned above and below the boron planes, introducing an intrinsic doping of the boron bands. The in-plane stretching of B-B bonds also gives rise to a phonon mode characterized by E$_{2g}$ symmetry. Due to the strong EPC, the E$_{2g}$ mode is responsible for a superconducting transition~\cite{bohnen01, kortus01, liu01, choi02, eiguren08, calandra10, margine13}. Furthermore, due to its significant anharmonicity, the strongly temperature-dependent Raman spectra can only be explained outside the scope of the standard anharmonic theory, by employing higher-order electron-phonon scattering within the framework of the NA theory~\cite{cappelluti06, saitta08, yildrim01, ponosov17}. The corresponding first-principles theoretical advances beyond the standard NA theory have already been made for EPC in certain cases~\cite{novko2018a,novko16,novko2018b,novko2020a}.
These amount to employing memory function methods originally derived to calculate intraband relaxation processes in the dynamical conductivity~\cite{marsiglio08,Kupcic2005,kupcic17,novko17,kupcic15,allen15,Sharapov05,Chakraborty78}, Raman response functions~\cite{kupcic08} or even exciton-phonon coupling~\cite{Antonius22}. In this way, one can account for the dynamical higher-order electron-phonon scattering, through the inclusion of a phonon-induced electron-hole pair self-energy by solving the Holstein problem~\cite{HOLSTEIN64}.

Here we perform a thorough quantitative analysis of the NA effects in prototypical 2D material graphene. This gives a basis for assessing the role of EPC in comprehending various phonon-related properties, such as thermal transport and superconductivity~\cite{profeta12, ludbrook15, ichinokura16, weller05, calandra05, margine13}, in layered and 2D compounds. We introduce the dynamical higher-order electron-phonon scatterings into the phonon self-energy, which gives us a tool to systematically explore the role of electron-mediated anharmonicity\,\cite{varma83,yoshiyama86,flicker15,novko2018b} in the phonon dynamics of graphene, and make a comparison with the well-known lattice-driven anharmonicity.
From first principles, we include these EPC effects through the temperature-dependent phonon-induced electron-hole pair self-energy, producing energy renormalization and finite lifetime. These two electron-phonon quantities introduce a crucial temperature dependence in the G mode Raman frequency and linewidth, showing anharmonic-like trends, contrary to previous works using a phenomenological broadening~\cite{giustino17,saitta08}. Through our systematic study, we show that the conventional and EPC-driven anharmonicites have different roles in temperature dependence of the G mode for undoped and doped graphene. In undoped graphene the lattice-driven anharmonicity (i.e., three-phonon and four-phonon scattering terms) is a dominant contribution to the G phonon linewidth, while for moderately-doped graphene (e.g., when $E_F=400$\,meV) both EPC and conventional anharmonicity are of the same order of magnitude. At the other extreme, for highly-doped graphene doped with alkali atoms\,\cite{howard11, torres12} or FeCl$_3$\,\cite{chen23}, the dominant role is played by the electron-mediated phonon-phonon coupling. Furthermore, we show that the latter interaction is the main relaxation channel for the G phonon in non-equilibrium graphene\,\cite{yan09,ishioka08,ferrante18,shengmeng22,girotto23}. We emphasize the generality of this fully first-principles approach.

\begin{figure*}[t]
\includegraphics[width=\textwidth]{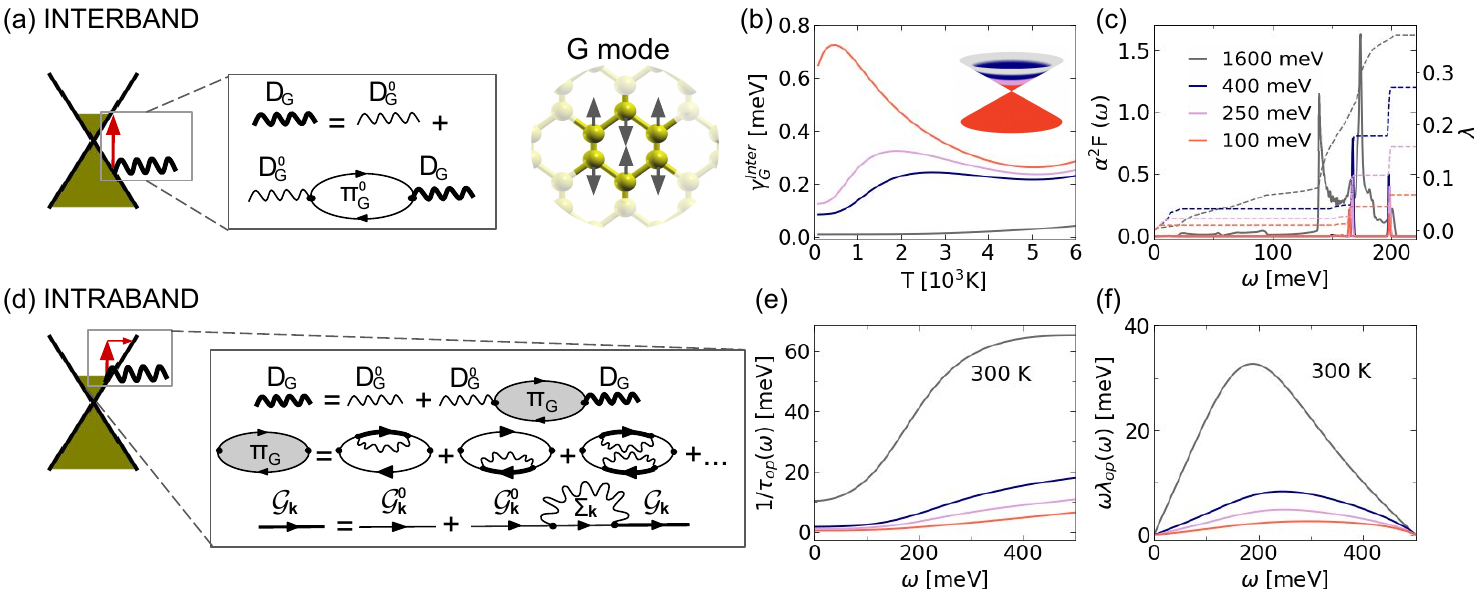}
\caption{\textbf{Renormalization of the G phonon mode in graphene due to electron-phonon coupling.} Two contributions to the $\mathbf{q}\simeq0$ phonon self-energy come from the (a) interband and (b) intraband electronic transitions. Interband transitions are directly facilitated by the schematically shown optical G mode. D$_{\text{G}}$ is the interacting propagator for the G mode, and D$^0_{\text{G}}$ is the bare one. The Dyson equation for the phonon propagator, in the first order of electron-phonon interaction in this case, contains only the interband phonon self-energy [$\Pi_G^0(\omega)$]. (b) The G phonon linewidth due to coupling to interband electron transitions for different dopings, which are schematically shown with a color-coded Dirac cone. (c) Adiabatic Eliashberg spectral function $\alpha^2F(\omega)$ for different dopings, showing an increase in the electron-phonon coupling constant $\lambda$ with doping. The cumulative $\lambda(\omega)$ for $E_F=100$\,meV, $E_F=250$\,meV, and $E_F=400$\,meV are multiplied by a factor of 10. (d) $\mathbf{q}\simeq0$ intraband transitions are a two-step process. Phonon excites an electron-hole pair, which in turn interacts with all the phonons in the system. However, the electron-hole pair is itself a quasiparticle, dressed by EPC. $\mathcal{G}_{\mathbf{k}}$ is the interacting electron (hole) propagator, and $\mathcal{G}^0_{\mathbf{k}}$ is the bare one. The symbol $\Sigma_{\mathbf{k}}$ stands for the electron self-energy. The Feynman diagrams for this process are shown in the inset of (d). (e) Inverse lifetime of an electron-hole pair $1/\tau_{\rm op}(\omega)$ at 300 K for different dopings, induced by interaction with phonons. (f) Energy renormalization of an electron-hole pair $\omega\lambda_{\rm op}(\omega)$ at 300 K for different dopings. }
\label{fig:fig1}
\end{figure*}

\section*{Results}
\textbf{Electron-mediated anharmonicity.} EPC renormalizes the bare phonon frequencies and introduces a finite lifetime to otherwise infinitely sharp phonons. These two effects are quantitatively contained within the phonon self-energy arising from EPC~\cite{giustino17} 
\begin{equation}
\pi_{\nu}(\mathbf{q},\omega)=\sum_{\mathbf{k}\text{nm}}\left| g_{\nu}^{\text{nm}}(\mathbf{k},\mathbf{q}) \right|^2\frac{f_{\text{n}\mathbf{k}+\mathbf{q}}-f_{\text{m}\mathbf{k}}}{\omega+\varepsilon_{\text{n}\mathbf{k}+\mathbf{q}}-\varepsilon_{\text{m}\mathbf{k}}+i\eta}.
\label{eq:eq1}
\end{equation}
The indices $n,m$ denote the electron band, $\mathbf{k}$ is the electron momentum and $f_{\text{m}\mathbf{k}} = 1/(e^{\beta(\varepsilon_{\text{m}\mathbf{k}}-\mu(\text{T}_e))}+1)$ is the Fermi-Dirac occupation factor at energy $\varepsilon_{\text{m}\mathbf{k}}$. Temperature dependence T$_e$ is contained within the factor $\beta = 1/(k_{B}\text{T}_e)$, where $k_{B}$ is the Boltzmann constant and is also indicated in the chemical potential $\mu$.
Electron-phonon matrix elements are denoted by $\left|g_{\nu}^{\text{nm}}(\mathbf{k},\mathbf{q}) \right|$, where $\mathbf{q}$ and $\nu$ are the phonon momentum and mode index. 
In density functional perturbation theory (DFPT)~\cite{baroni01}, phonons are statically screened by the electron gas, which is why one needs to renormalize them with NA corrections~\cite{giustino17,nina23,novko16}. The real part of the phonon self-energy shifts the adiabatic (DFPT) frequencies $\omega_{\nu\mathbf{q}}^{\rm NA} =\omega_{\nu\mathbf{q}}^{\rm A} +  \text{Re}[\pi_{\nu}(\mathbf{q},\omega^{\rm NA})] -  \text{Re}[\pi_{\nu}(\mathbf{q},0)]$, while the imaginary part  introduces the broadening $\gamma_{\nu\mathbf{q}}= -2\text{Im}[\pi_{\nu}(\mathbf{q},\omega^{\rm NA})]$. In order to describe the Raman experiments, where light has vanishing momentum, we need to consider the long-wavelength limit ($\mathbf{q}\simeq 0$) in Eq.~\eqref{eq:eq1}. If the static part of the phonon self-energy, already accounted for in DFPT, is subtracted from Eq.~\eqref{eq:eq1}, we obtain two first-order contributions to the phonon renormalization. The first one corresponds to the adiabatic intraband 
\begin{equation}
\pi_{\nu}^{\text{intra}}(0,0)=\sum_{\mathbf{k}\text{n}}\left| g_{\nu}^{\text{n}}(\mathbf{k},0) \right|^2\left(-\frac{\partial f_{\text{n}\mathbf{k}}}{\partial \varepsilon_{\text{n}\mathbf{k}}}\right)
\label{eq:eq2}
\end{equation}
and the second one to the dynamical interband electron transitions:
\begin{equation}
\pi_{\nu}^{\text{inter}}(0,\omega)=\sum_{\mathbf{k}\text{n}\neq \text{m}}\left| g_{\nu}^{\text{nm}}(\mathbf{k},0) \right|^2\frac{f_{\text{n}\mathbf{k}}-f_{\text{m}\mathbf{k}}}{\omega+\varepsilon_{\text{n}\mathbf{k}}-\varepsilon_{\text{m}\mathbf{k}}+i\eta},
\label{eq:eq3}
\end{equation}
where $\eta$ is the infinitesimal parameter.
Within first-order perturbation theory, NA effects can be included only in the interband channel, since only that term has a frequency dependence. In Fig.~\ref{fig:fig1}(a), we schematically show the vertical ($\mathbf{q}\simeq 0$) interband transitions induced by the optical G mode with energy of  $\sim 200$\,meV. The phonon linewidth [see Fig.~\ref{fig:fig1}(b)] deriving from these contributions can be written as
\begin{equation}
\gamma^{\rm inter}_{\nu}( 0,\omega)=-2\eta\sum_{\mathbf{k}\text{nm}}\left| g_{\nu}^{\text{nm}}(\mathbf{k},0) \right|^2\frac{f_{\text{n}\mathbf{k}}-f_{\text{m}\mathbf{k}}}{(\omega+\varepsilon_{\text{n}\mathbf{k}}-\varepsilon_{\text{m}\mathbf{k}})^2+\eta^2},
\label{eq:inter_imag}
\end{equation}
and it reduces as the doping or temperature is increased due to Pauli blocking. When analyzing the Raman spectra of graphene, this interband EPC contribution to the linewidth is commonly considered as the only EPC contribution~\cite{Han22,bonini07}, which is why EPC is considered to provide negligible contributions at higher temperatures or dopings. 

In Fig.~\ref{fig:fig1}(c) we show the adiabatic Eliashberg spectral function for different dopings in graphene. The Eliashberg spectral function reveals the energy distribution of strongly coupled phonons and is defined as 
\begin{equation}
\alpha^2F(\omega)=\frac{1}{\pi \text{N$_0$}}\sum_{\mathbf{q}\nu}\frac{\gamma_{\mathbf{q}\nu}}{\omega_{\mathbf{q}\nu}}\delta(\omega_{\mathbf{q}\nu}-\omega),
\label{eq:a2f}
\end{equation}
with $\text{N$_0$}$ being the electronic density of states (DOS) at the chemical potential. In the same plot, we show a cumulative EPC constant $\lambda$ showing an increase in EPC with doping. The main contributions come from the G mode and A$'_1$ mode at the K point of the Brillouin zone. The cumulative $\lambda(\omega)$ for $E_F=100$\,meV, $E_F=250$\,meV, and $E_F=400$\,meV are multiplied by a factor of 10, so that the doping-induced increase is more evident. 

The first-order intraband term is static and contains scattering processes within a narrow energy window around the chemical potential, broadened by temperature [see Eq.~\eqref{eq:eq2}]. It has no imaginary part, so it does not contribute to the EPC-induced linewidth. In graphene and other materials, even though the dynamical first-order intraband phonon self-energy contributions vanish in the long-wavelength limit, Raman experiments still suggest that EPC in fact must somehow contribute to the linewidth. In the first-order perturbation theory, the excited phonon interacts with electrons via the generation of ``bare'' electron-hole pairs, while the processes where the generated electron-hole pairs are renormalized by the rest of the system's phonons are discarded. 

The latter multi-step process is schematically shown in Fig.~\ref{fig:fig1}(d). The Dyson equation in the inset of Fig.~\ref{fig:fig1}(d) shows how the phonon self-energy bubble is renormalized by diagrams describing electron-hole scattering with all the phonons of the system. Essentially, electrons mediate phonon-phonon interaction\,\cite{varma83,yoshiyama86,flicker15,novko2018b}. Temperature effects can strongly alter the distribution of electronic excitations~\cite{ponosov94,cerdeira72} and the bare electronic structure and electron DOS (N$_0$) can be significantly renormalized by phonons if it varies on their energy scale~\cite{Knigavko2005}. 
This subtlety is denoted in the Dyson equation in Fig.~\ref{fig:fig1}(d) by \textit{dressed} electron (hole) propagator lines. The self-consistent self-energy correction leads to higher-order EPC contributions and the appearance of the rainbow diagrams\,\cite{Bianco2023NonperturbativeTO}.
Diagrams in Fig.~\ref{fig:fig1}(d) can be summed up in a Bethe-Salpeter equation for the full phonon self-energy, which can be schematically written as
\begin{equation}
    \Pi = \Pi^0 + \Pi^0 M \Pi 
    \label{eq:pi_bse}
\end{equation}
with all the wavevector and energy subscripts omitted. $M(\omega)$ is the electron-hole self-energy and can be written as a sum of its real and imaginary parts
\begin{equation}
    M (\omega,T)  = \frac{i}{\tau_{\rm op}(\omega,T)} + \omega\lambda_{\rm op}(\omega,T),
\end{equation}
where the wavevector indices are omitted and temperature dependence is indicated. 
Considering electron-phonon scattering processes up to all orders and solving the Bethe-Salpeter equation~\eqref{eq:pi_bse}, leads to the following form for the intraband phonon self-energy in the long-wavelength limit~\cite{novko16,novko2018a}:
\begin{equation}
\begin{aligned}
    \pi_{\nu}^{\text{intra}}(0,\omega)=\sum_{\mathbf{k}\text{n}}\left|g_{\nu}^{\text{n}}(\mathbf{k},0)\right|^2\left(-\frac{\partial f_{\text{n}\mathbf{k}}}{\partial \varepsilon_{\text{n}\mathbf{k}}}\right) \\
\frac{\omega}{\omega[1+\lambda_{\rm op}(\omega)] + i/\tau_{\rm op}(\omega)}.
\label{eq:eq4}
\end{aligned}
\end{equation}
Instead of a vanishing dynamical contribution obtained through first-order perturbation theory, we obtain a finite contribution with an additional temperature-dependent factor deriving from the complex electron-hole self-energy $M(\omega,T)$. An equivalent expression can be obtained by means of Green's functions~\cite{carbotte92,maksimov96}. 
Throughout the paper we will refer to expressions Eqs.~\eqref{eq:eq3} and \eqref{eq:eq4} as interband and intraband phonon self-energies, respectively.

\begin{figure}[!t]
\includegraphics[width=0.5\textwidth]{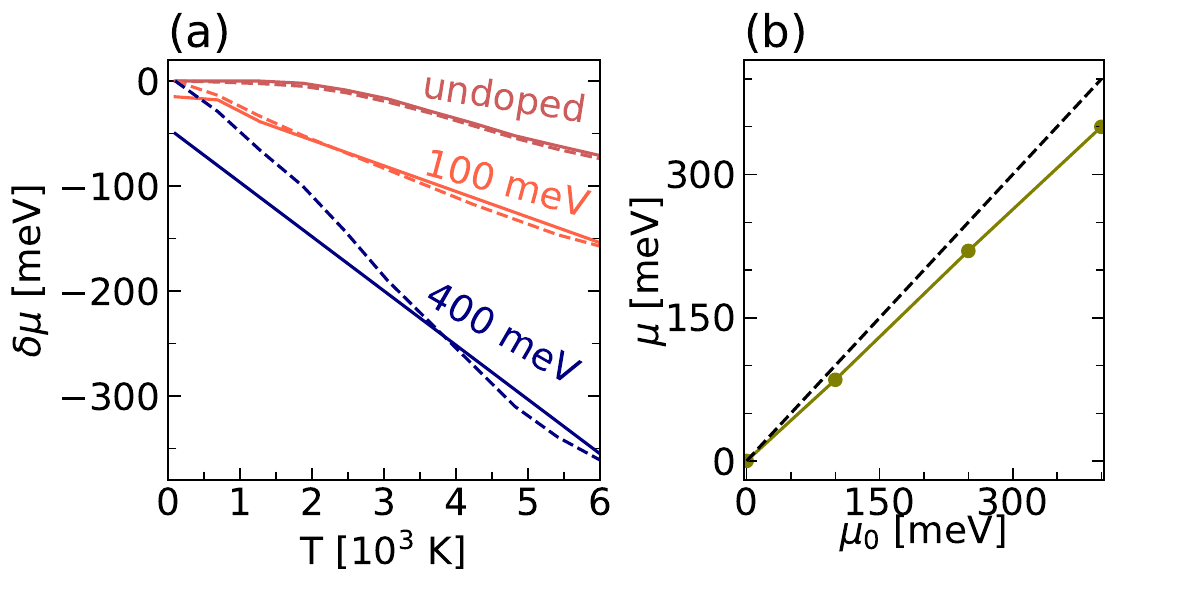}
\caption{\textbf{The conservation of number of carriers.} (a) Temperature-dependent chemical potential for various doping levels in graphene. Dashed line denotes the bare result, while a self-consistent result is denoted with a full line. (b) T=0 EPC induced Fermi level renormalization for various doping levels in graphene. Dashed line corresponds to an unrenormalized value. 
} 
\label{fig:fig1a}
\end{figure}

An interacting electron-hole pair has a temperature-dependent finite inverse lifetime $1/\tau_{\rm op}(\omega)$ [see Fig.~\ref{fig:fig1}(e)] and energy shift $\omega\lambda_{\rm op}(\omega)$ [Fig.~\ref{fig:fig1}(f)]. Unlike the interband linewidth contributions, these two quantities increase with the doping level. One can obtain the electron-hole self-energy from optical conductivity measurements where it enters the extended Drude optical conductivity~\cite{puchkov96}. For this reason, $1/\tau_{\rm op}(\omega)$ is sometimes called optical electron-hole scattering rate. It can then be used to obtain the electron-phonon spectral function~\cite{Carbotte99,allen71,marsiglio98}. Theoretically, it is more convenient to go in the opposite direction and calculate the electron-hole self-energy at finite temperatures from the Eliashberg spectral function as~\cite{shulga91,novko2018a,novko2020a}
\begin{dmath}
\frac{1}{\tau_{\rm op}(\omega)}= \frac{\pi}{\omega} \int_{0}^{\infty}d\Omega \alpha^2F(\Omega)\left[2\omega \text{coth}\frac{\Omega}{2{\rm T}_{\rm ph}}-(\omega + \Omega)\text{coth}\frac{\Omega+\omega}{2\text{T}_e} +(\omega - \Omega)\text{coth}\frac{\omega-\Omega}{2\text{T}_e } \right].
\label{eq:inv_tau}
\end{dmath}
$\text{T}_{\rm ph}$ denotes the phonon temperature, while with $\text{T}_e$ we denote the temperature of the electronic system. In graphene, the calculations are performed for its G mode, and the label $\text{T}_{\rm ph}$ is exchanged with $\text{T}_{G}$. In our work, we use a further extension to include the effects of electron band structure, which were already discussed for normal and superconducting state properties in Ref.~\citenum{pickett80,Pickett82,Mitrovic83}. 
The zero temperature result for the electron-hole self-energy derived in Ref.~\citenum{mitrovic85} was extended to finite temperatures in Ref.~\citenum{Sharapov05} and the resulting expression is
\begin{dmath}
\frac{1}{\tau_{\rm op}(\omega)}= \frac{\pi}{\omega} \int_{0}^{\infty}d\Omega \alpha^2F(\Omega)\newline \int_{-\infty}^{\infty}d\varepsilon\left\{\frac{\Tilde{N}(\varepsilon-\Omega)}{N_0} + \frac{\Tilde{N}(-\varepsilon+\Omega)}{N_0}\right\} \newline [n_{B}(\Omega) + f(\Omega-\varepsilon)][f(\varepsilon-\omega)-f(\varepsilon+\omega)].
\label{eq:inv_tau2}
\end{dmath}
The factors in Eq.~\eqref{eq:inv_tau2}, include the Eliashberg spectral function, the renormalized or quasi-particle DOS $\Tilde{N}(\varepsilon)$ as well as the Fermi-Dirac f($\varepsilon$) and Bose-Einstein n$_B(\omega$) occupation factors. The same expression was used for description of higher electron-phonon scattering terms in optical conductivity formula\,\cite{Sharapov05}. The real part of the electron-hole self-energy $\omega\lambda_{\rm op}(\omega)$ can be obtained with the Kramers-Kronig transformation of $1/\tau_{\rm op}(\omega)$.
The quasi-particle DOS is calculated as 
\begin{dmath}
\Tilde{N}(\varepsilon)=\int d\mathbf{k} A(\varepsilon,\mathbf{k}),
\label{eq:N_tilda}
\end{dmath}
which needs to be solved self-consistently in combination with the equation for the electron spectral function 
(imaginary part of the electron propagator $\mathcal{G}_{\mathbf{k}}$)
\begin{dmath}
A(\varepsilon,\mathbf{k})=\frac{1}{\pi}\frac{-\text{Im}\Sigma(\varepsilon)}{[\varepsilon-\varepsilon_{\mathbf{k}}+\mu-\text{Re}\Sigma(\varepsilon)]^2 + [\text{Im}\Sigma(\varepsilon)]^2},
\label{eq:A}
\end{dmath}
reachable in angle-resolved photoemission spectroscopy, 
and the imaginary part of the electron self-energy 
\begin{dmath}
\text{Im}\Sigma(\varepsilon) = -\pi \int_{0}^{\infty} d\Omega \alpha^2F(\Omega) \left\{ \frac{\Tilde{N}(\varepsilon-\Omega)}{N_0} [ n_{B}(\Omega) + 1 + f(\varepsilon-\Omega)] + \frac{\Tilde{N}(\varepsilon+\Omega)}{N_0} [ n_{B}(\Omega) + f(\varepsilon+\Omega)]\right\}.
\label{eq:ImSigma}
\end{dmath}

\begin{figure}[!t]
\includegraphics[width=0.5\textwidth]{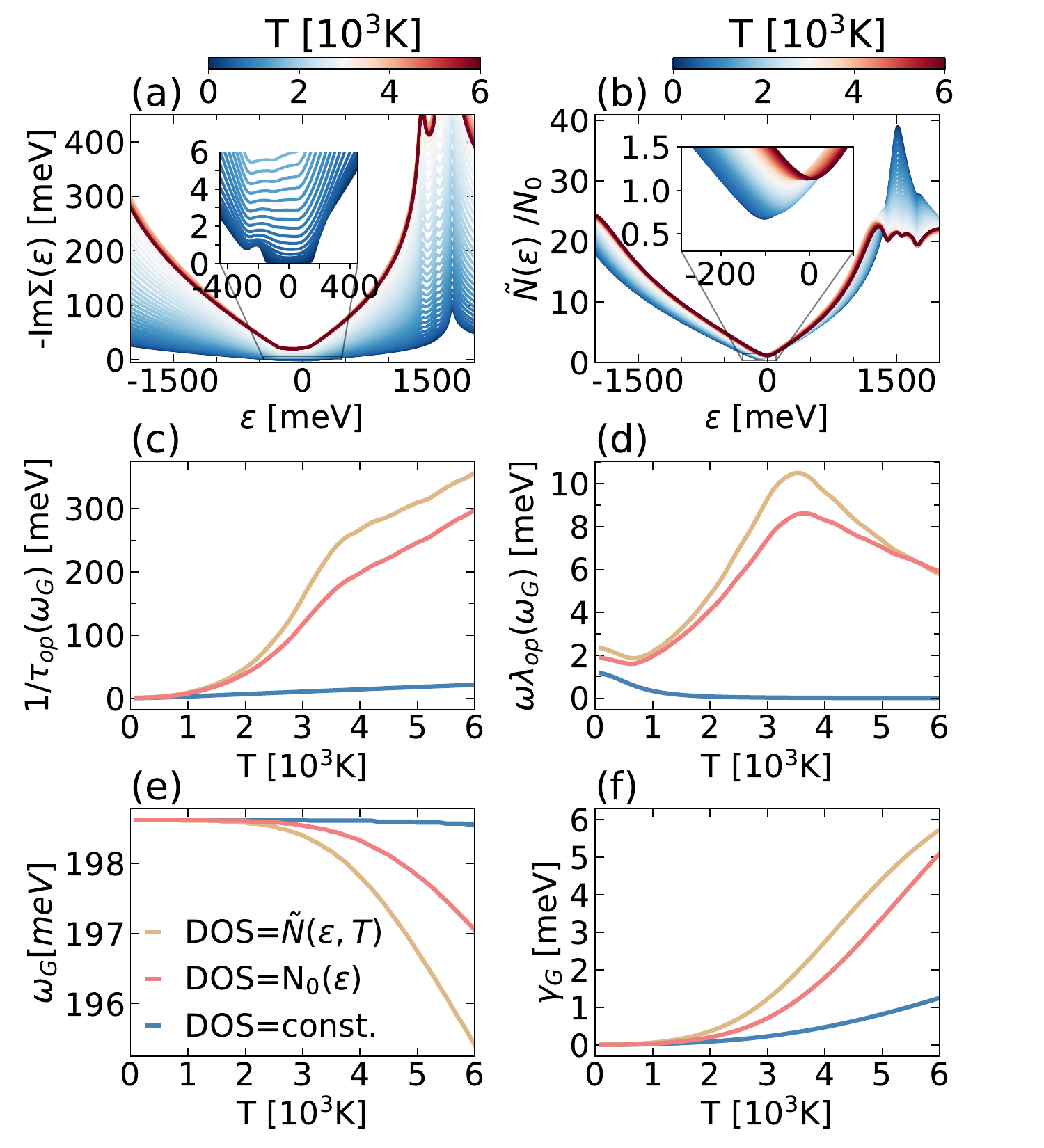}
\caption{\textbf{Temperature-dependent one- and two-particle properties of graphene with $\mathbf{E_{F}=100}$\,meV.} (a) Imaginary part of the electron self-energy $\mathrm{Im}\,\Sigma(\varepsilon)$. (b) Quasi-particle DOS $\tilde{N}(\varepsilon)/N_0$. (c)-(f) Comparison of three different approaches [constant DOS, bare DOS $N_0(\varepsilon)$ and quasi-particle DOS $\tilde{N}(\varepsilon)$] in the calculation of the temperature dependence of (c) electron-hole pair inverse lifetime $1/\tau_{\rm op}(\omega_G)$, and (d) energy renormalization $\omega\lambda_{\rm op}(\omega_G)$, as well as the corresponding (e) G mode frequency renormalization $\omega_G$ and (f) G mode linewidth $\gamma_G$.
} 
\label{fig:fig2}
\end{figure}

The energy and temperature dependence in the quasi-particle DOS $\Tilde{N}(\varepsilon)$ then additionally affects the inverse lifetime and EPC constant~\cite{Mitrovic83}. Note that the inclusion of the quasi-particle DOS in the calculation of the electron self-energy is equivalent to using interacting momentum distribution functions instead of the bare Fermi-Dirac distribution~\cite{kupcic15}. 
Along with the quasiparticle DOS $\Tilde{N}(\varepsilon)$ and spectral function $A(\varepsilon,\mathbf{k})$, in the self-consistent calculation one also needs to include the chemical potential shift with temperature. The chemical potential fixes of the number of carriers $n_0$, using the renormalized DOS. As temperature increases or as interactions are turned on (or both)\,\cite{luttinger60}, the chemical potential needs to be adjusted in order to conserve the number of electrons  
\begin{dmath}
    n_0 = \int^{\infty}_{-\infty}d\varepsilon N_0(\varepsilon)f(\varepsilon,0)  \\ = \int^{\infty}_{-\infty}d\varepsilon \Tilde{N}(\varepsilon+\delta\mu(0))f(\varepsilon,0) = \int^{\infty}_{-\infty}d\varepsilon \Tilde{N}(\varepsilon+\delta\mu(T))f(\varepsilon,T).
    \label{eq:n_qp}
\end{dmath}
where energies $\varepsilon$ are measured from the T=0 Fermi level. Equations~\eqref{eq:N_tilda}-\eqref{eq:n_qp} need to be iterated self-consistently for every temperature until the temperature-dependent chemical potential and the corresponding quasi-particle DOS conserve the number of particles. Due to the complexity of the procedure, when calculating the chemical potential shifts, we resort to the \textit{on-shell} approximation in the electron self-energy Eq.~\eqref{eq:ImSigma} where the general energy argument $\varepsilon$ is substituted by the Kohn-Sham value $\varepsilon_{n\mathbf{k}}$\,\cite{giustino17}. Note that the new value for the chemical potential enters the $\Tilde{N}(\varepsilon)$ indirectly through the spectral function and the $\text{N$_0$}$. 

In Fig.~\ref{fig:fig1a} we show the result of a self-consistent chemical potential calculation. The bare chemical potential temperature shift remains smaller than the self-consistent result until some finite temperature value, when the trend reverses [see Fig.~\ref{fig:fig1a} (a)]. At the highest considered temperature, the bare temperature and interaction induced chemical potential shift is larger than the shift obtained in a self-consistent calculation. Even though the observed differences between the two are quite small, we find that interactions mitigate the chemical potential temperature shift. In agreement with Ref.~\citenum{Carbotte10}, we find that in comparison with the $\mu(T\rightarrow$0) calculation from the bare DOS, the self-consistent calculation leads to an interaction-induced reduction of the chemical potential [see Fig.~\ref{fig:fig1a} (b)].

In Fig.~\ref{fig:fig2}(a) we show the temperature variations in the imaginary part of the electron self-energy. At low temperatures, a characteristic energy dependence can be recognized resulting from the linear bands and the interaction with the optical phonons (i.e., G and A$_1'$ modes)~\cite{Carbotte10,park07}. The resulting renormalization of the quasi-particle DOS with temperature is shown in Fig.~\ref{fig:fig2}(b). The remaining panels in Fig.~\ref{fig:fig2} show how the three different approaches to the $1/\tau_{\rm op}(\omega)$ calculation affect the electron-hole and phonon properties. The simplest approximation amounts to neglecting the energy dependence in the DOS. For graphene, it leads to almost no renormalization of the $\omega_G$ frequency, and weak linewidth contribution (blue line). On the other hand, considering the variations of the DOS with energy leads to larger values for the electron-hole self-energy, strong G mode renormalizations and significant linewidth contribution (red and yellow lines). The characteristic rapidly increasing DOS in graphene with $E_{F}=100$\,meV, leads to the higher contribution coming from the ratio $N_0(\varepsilon)/N_0$ in Eq.~\eqref{eq:inv_tau2} (red), or $\Tilde{N}(\varepsilon)/N_0$ for the case when quasi-particle DOS is included (yellow), which is then almost always larger than 1. Note how the proper inclusion of dynamical EPC leads to a quadratic temperature dependence of $1/\tau_{\rm op}(\omega)$ in the low temperature limit~\cite{allen74,nina23}. Here we emphasize once again that to obtain the proper EPC-induced temperature-dependent phonon dynamics, it is crucial to account for the full self-consistent interactions of electron-hole pairs (inclusion of electron-hole self-energies with renormalized quasi-particle electronic structure) and conservation of the number of carriers (inclusion of interacting chemical potential shifts), both of which are usually omitted in the simulation of the phonon dynamics of Raman-active modes in graphene, as well as in other systems where EPC plays an important role. Using the model Hamiltonians, this electron-mediated anharmonicity was argued to be important for charge-density-wave (CDW) materials where EPC is strong and these higher-order effects contribute to the CDW transition temperature and renormalization of electronic structure\,\cite{varma83,flicker15,yoshiyama86}.

Diagrams with phonons connecting the electron and hole lines inside the bubble are called vertex corrections, and are omitted in this derivation.  
It could be argued that neglecting vertex corrections introduces a degree of uncertainty, but all first-principles calculations to date are subject to this limitation. Also, conflicting evidence in the literature regarding the validity of Migdal's theorem~\cite{migdal58} ranges from its breakdown in the NA limit, and the possibility that vertex corrections wash out the NA effects~\cite{grimaldi95b,pietronero96,itai92}, to evidence that the vertex corrections lead to insignificant qualitative changes~\cite{maksimov96}, even for large couplings~\cite{gunnarsson12}. There is also additional evidence of its validity in (doped) graphene~\cite{dassarma14}. The inclusion of the vertex corrections is non-trivial and is left for future work.

Further, it is important to stress that we use a standard semi-local DFT functional in order to calculate the EPC properties for graphene, while it was shown that beyond-DFT corrections to electron correlations (e.g., GW approximation and hybrid functionals) modify the electronic structure, increase the Fermi velocity around the Dirac points\,\cite{Trevisanutto2008}, and enhance the EPC strengths of both G and A$_1'$ phonon modes\,\cite{lazzeri08,attaccalite10}. These beyond-DFT corrections might impact the evaluation of the intraband phonon self-energy, however, it is not clear to what extent, considering that the increase of the Fermi velocity will decrease the value of $\partial f / \partial \varepsilon$ and $1/\tau_{\rm op}$ in Eq.\,\eqref{eq:eq4}, while the increase of the EPC strength will increase $g^2$ and $1/\tau_{\rm op}$. The competition of these electron correlation effects in the evaluation of the phonon self-energy is an interesting problem that we also leave for the future work.

\textbf{Conventional anharmonicity.} The ionic motion is determined by the Born-Oppenheimer
potential, which is usually expanded up to second order in ionic displacements, leading to a parabolic harmonic potential well. However, in some materials or at high temperature, deviations from the harmonic model can not be neglected, and the parabolic potential landscape will be modified both in shape and in its equilibrium position.

If anharmonicity cannot be addressed perturbatively, it becomes challenging to relinquish the benefits of treating phonons as well-defined quasiparticles. Several methods have already been developed to treat this complexity, based on molecular dynamics (MD)~\cite{car1985} (or path-integral MD if quantum effects in ionic motion are included~\cite{Ceperley1995}), or on a stochastic sampling of the potential energy surface. We mostly focus on the latter in our analysis of graphene's Raman frequencies and linewidths.

The Temperature Dependent Effective Potential (TDEP) method extracts interatomic force constants (IFCs) by fitting displacements and forces directly from  molecular dynamics (MD) simulations~\cite{Knoop2024,Castellano2023,Hellman2011,Hellman2013,Hellman2013b} driven by forces from ab initio or Machine Learning (ML) calculations. This method intrinsically incorporates anharmonic effects both at the harmonic level, giving rise to an effective harmonic theory, and at higher orders (presently 3rd and 4th order IFCs), all of which are now explicitly temperature dependent. 

The stochastic self-consistent harmonic approximation (SSCHA \cite{Monacelli2021, errea2014, bianco2017}), on the other hand, is based on the Gibbs–Bogoliubov free energy variational principle. Within SSCHA, the Gibbs-Bogoliubov free energy functional is minimized by employing trial harmonic potentials, and a stochastic approach is used to sample the potential energy surface according to the trial harmonic statistics. SSCHA inherently accounts for both anharmonicity—incorporating contributions from all even orders of the potential energy surface—and quantum effects, as nuclear dynamics are fully treated at the quantum level, with the nuclear kinetic operator explicitly taken into account. However, the stochastic ensembles are constrained to be Gaussian. Once the minimization is performed and the free energy is obtained, an effective set of second-order interatomic force constants (IFCs) naturally emerges, defining the stochastic self-consistent harmonic (SCHA) noninteracting quasiparticles. The TDEP code has implemented an equivalent method using stochastic sampling, but with its own IFC fitting approach. This method is commonly known as sTDEP \cite{shulumba17} and in this work we compare it with the SSCHA method.

\begin{figure}[!t]
\includegraphics[width=0.5\textwidth]{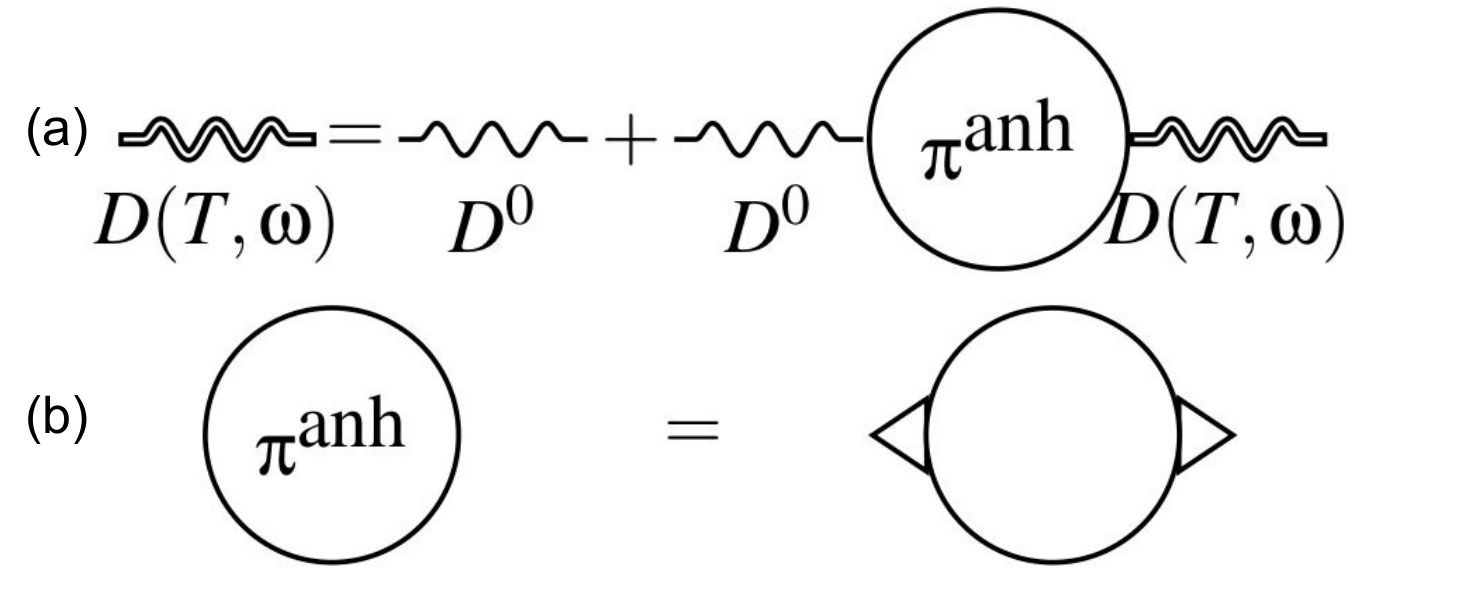}
\caption{\textbf{Conventional anharmonic contributions.} (a) Diagrammatic representation of the Dyson equation for the anharmonic phonon propagator. (b) The anharmonic phonon self-energy diagram.
} 
\label{fig:fig3a}
\end{figure}

The Dyson equation describing the evolution of the interacting anharmonic phonon propagator can be written as 
\begin{equation}
    D^{\text{anh}}(\mathbf{q},\omega) = D^{\text{0}}(\mathbf{q},\omega) + D^{\text{0}}(\mathbf{q},\omega) \pi^{\text{anh}}(\mathbf{q},\omega)D^{\text{anh}}(\mathbf{q},\omega),
\end{equation}
where $\pi^{\text{anh}}(\mathbf{q},\omega)$ is the self-energy of the non-interacting anharmonic (e.g., SCHA) phonons. The Dyson equation is diagrammatically written in Fig.~\ref{fig:fig3a}(a). 
The single wavy line denotes a renormalized but non interacting effective \emph{harmonic} phonon, while the double wavy line includes the effects of anharmonic interactions contained in the residual anharmonic phonon self-energy. 

Using the SSCHA code, we calculated the SCHA-phonon self-energy, explicitly accounting for interactions between SCHA phonons. We restricted our analysis to the so-called bubble approximation, shown diagrammatically in Fig.~\ref{fig:fig3a}(b), where only third-order SCHA interaction vertices are included. These vertices are defined as the stochastic averages of the third-order derivatives of the potential energy surface with respect to atomic positions. In our evaluation of the SCHA-phonon self-energy, we did not include fourth-order SCHA interaction vertices~\cite{Monacelli2021}. Within the TDEP framework, we computed a similar phonon self-energy but including both third- and fourth-order interaction vertices between the auxiliary quasiparticles.

\begin{figure*}[!t]
\includegraphics[width=\textwidth]{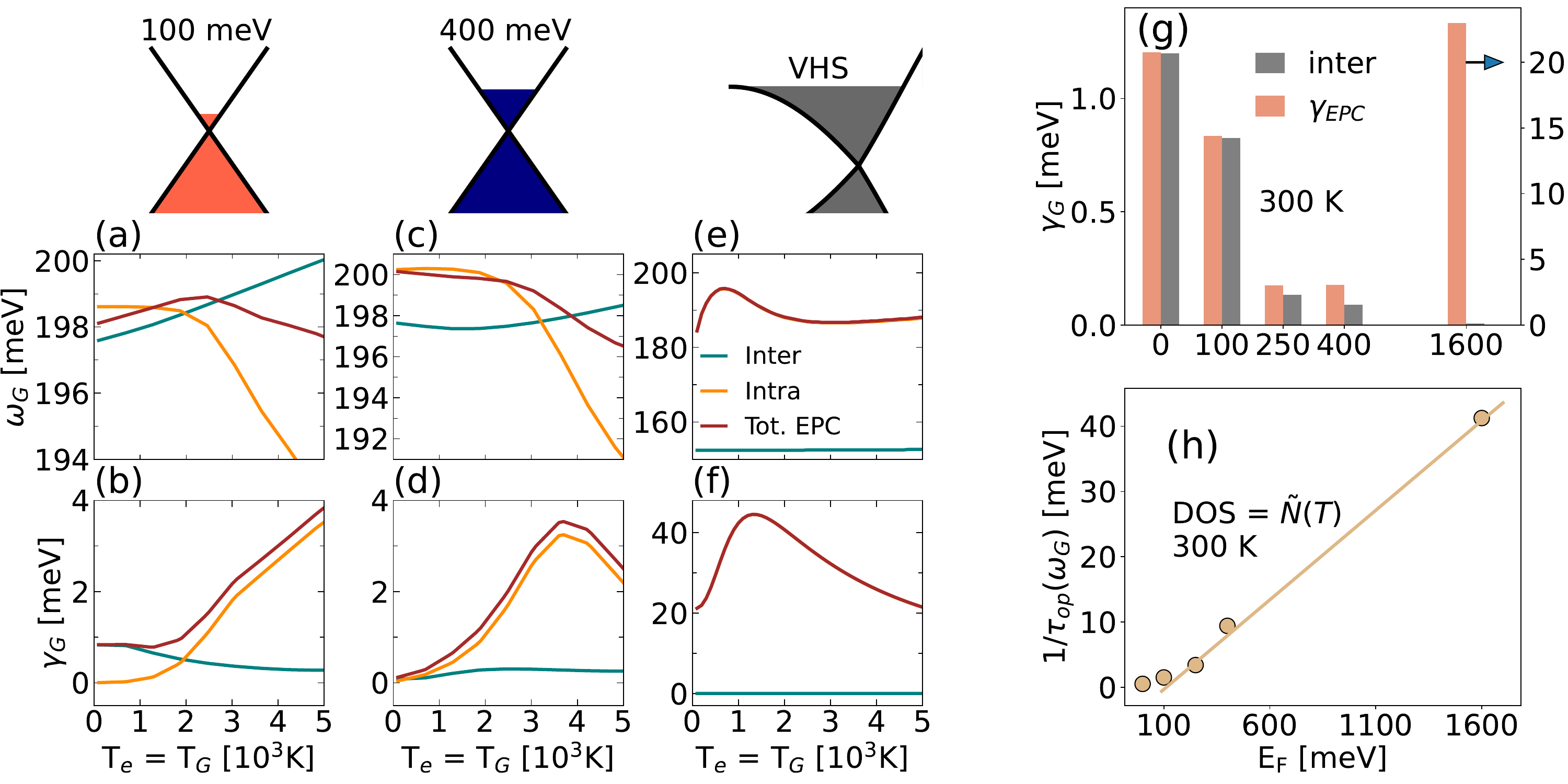}
\caption{\textbf{The role of electron-mediated anharmonicity in doped graphene.} For the three doping regimes, we show the total G mode frequency $\omega_G$ renormalization in (a),(c),(e); and  linewidth $\gamma_G$ induced by EPC in (b),(d),(f). The intraband phonon self-energy contribution leads to the observed trend of (a) frequency decrease with temperature and (b) linewidth increase with temperature. (c) For larger doping we observe an even larger importance of the intraband contribution representing electron-mediated phonon-phonon coupling. For graphene doped up to the VHS, the intraband contribution is the dominant contribution, and leads to a large non-monotonic (e) frequency and (f) linewidth temperature dependence. Note that in panels (e) and (f) the intraband contribution (yellow curve) is hidden behind the the total result (red curve). (g) Comparison between the first-order interband (grey) and total contribution (red) to the linewidth $\gamma_G$ at 300 K for various dopings ($x$-axis in meV). The right $y$-axis is for the case of highest doping (VHS), while the left $y$-axis is for the rest. (h) Doping-dependent electron-hole pair inverse lifetime $1/\tau_{\rm op}(\omega_G)$ at 300\,K. 
} 
\label{fig:fig3}
\end{figure*}

The phonon spectral function is determined as the imaginary part of the phonon propagator. In this work, we compute it using the no-mode mixing approximation and further approximate it with a Lorentzian profile for each mode. We carry out calculations across a range of temperatures, then obtain the $\mathrm{T}=0$\,K frequency either by linearly extrapolating the temperature-dependent results or via molecular dynamics, excluding the zero-point frequency shift. This $\mathrm{T}=0$\,K frequency serves as a reference, enabling us to calculate the anharmonic phonon self-energy based on the frequency shift observed with temperature.

The full anharmonic propagator containing the EPC effects is then obtained as 
\begin{equation}
        B_{\nu}(\mathbf{q},\omega)=-\frac{1}{\pi}\mathrm{Im}\left[ \frac{2\omega^A_{\mathbf{q}\nu}}{\omega^2-(\omega^{\text{A}}_{\mathbf{q}\nu})^2-2\omega^{\text{A}}_{\mathbf{q}\nu}\pi^{\text{TOT.}}_{\nu}(\mathbf{q},\omega)} \right]
\label{eq:ch4_spec}
\end{equation}
where $\pi^{\text{TOT.}}_{\nu}(\mathbf{q},\omega)$ refers to the total phonon self-energy $\pi^{\text{NA}}_{\nu}(\mathbf{q},\omega)+\pi^{\text{anh}}_{\nu}(\mathbf{q},\omega)$.
The first self-energy term $\pi^{\text{NA}}_{\nu}(\mathbf{q},\omega)$ is the NA EPC-induced self-energy term, obtained by adding the dynamical interband and intraband contributions and subtracting the static contribution, already accounted for in $\omega^{\text{A}}_{\mathbf{q}\nu}$. The on-shell anharmonic phonon self-energy, from TDEP or SSCHA is denoted by $\pi^{\text{anh}}_{\nu}(\mathbf{q},\omega_{\mathbf{q}\nu})$. 

The G mode frequency is found as the position of the peak of the calculated phonon spectral function Eq.~\eqref{eq:ch4_spec}. Depending on which self-energy contributions are included, we obtain the interband (i.e., first-order EPC), intraband (i.e., higher-order EPC), anharmonic, and total phonon renormalization with temperature. Note that our final SSCHA and sTDEP results also incorporate the effects of the thermal expansion of the unit cell calculated using a quantum mechanical approach. The DFPT calculations and EPC properties do not include the effect of volume expansion.   

It is crucial to stress that with this approach we update the harmonic frequencies originally contained in $\pi^{\text{NA}}_{\nu}(\mathbf{q},\omega)$, but only partially. Namely, in the calculations of $\alpha^2 F(\omega)$ and $1/\tau_{\rm op} (\omega)$ we still keep the original harmonic phonon frequencies, considering that the overall (full Brillouin zone) phonon density of states in graphene is not significantly modified by the inclusion of anharmonic corrections. As a result, we do not expect important modifications to the total momentum-integrated EPC strength. Namely, from Supplementary Figure 1, where we compare the full dispersion of harmonic and SSCHA phonons, it is clear that strongly-coupled G and $A_1'$ phonons are modified only up to $\sim 5\%$ for a relevant range of temperatures.

\textbf{Raman spectrum of graphene.} In Fig.~\ref{fig:fig3} we show the role of EPC in G mode Raman features for three different dopings in successive columns. If the Fermi level is 100 meV above the Dirac point, a G phonon with energy of ~$200$\,meV directly facilitates the interband contribution to the EPC-induced phonon linewidth. In the second column, doping sets $E_F=400$\,meV, where Pauli blocking should lead to negligible interband contributions. The third case corresponds to graphene doped up to the Van Hove singularity (VHS) in the DOS, which should approximately correspond to highly-doped intercalated graphene\,\cite{howard11,chen23}. For the temperature dependence of both G mode frequency and linewidth we explicitly distinguish between the interband and intraband contributions to the phonon self-energy.
Focusing first on the G mode frequency shift [see Figs.~\ref{fig:fig3}(a),~\ref{fig:fig3}(c),~\ref{fig:fig3}(e)] we notice that the intraband contribution grows with increased doping or higher temperatures.
The interband contribution increases the G mode frequency with temperature until, for larger doping, its effect becomes negligible. 
Anharmonic effects are known to reduce the G mode frequency with an increase in temperature, especially the four-phonon contribution~\cite{lin11,bonini07,chen21}. Since the intraband contribution corresponds effectively to an electron-mediated phonon-phonon interaction\,\cite{varma83,novko2018b}, it shows a characteristic anharmonic-like increase with temperature. More striking contributions from the intraband phonon self-energy are observed for the G mode linewidth. Contrary to the common belief that the total EPC-induced linewidth comes from the interband phonon self-energy, we obtain a much larger intraband contribution, which has an opposite temperature trend. The weak initial linewidth decrease at low T, observed e.g. in Ref.~\citenum{chae10,liu19,giura12} comes from the EPC interband term and is therefore present only when the chemical potential is smaller than the half of the G mode frequency. The intraband contribution, on the other hand, increases with both T and doping. The total EPC-induced linewidth eventually turns out to be a solely intraband effect for moderately and highly-doped graphene.

Overlooking higher-order electron-phonon scattering effects leads to an incomplete description of the Raman G mode spectral features. The EPC-induced changes of frequency and linewidth with temperature are almost entirely attributable to the commonly disregarded intraband contribution, and the experimentally observed behavior results from a cooperative interplay with anharmonicity. This supports the conclusion drawn in Ref.~\citenum{Berciaud2010} that anharmonicity alone is not substantial enough to account for the full linewidth. Interestingly, our results for the highest doping case resembles the Raman spectroscopy experimental result from Ref.~\citenum{Tiberj2013}, where photo-doping is induced with lasers of high power. According to Berciaud~\cite{Berciaud2010} this is equivalent to large temperatures with $\text{T}_e=\text{T}_G$. They report an unconventional decrease, followed by an increase in frequency, and an increase followed by a decrease in linewidth as a function of laser power (i.e., increase of temperature). The (rigid band) doping level used in our work, differs from the experimental one from Ref.~\citenum{Tiberj2013}, but the frequency and linewidth temperature dependence show very similar non-monotonous trends. The only mechanism able to explain such behaviour is higher-order EPC: the standard lattice-driven anharmonic contribution is constantly increasing. The same unconventional trend in the temperature dependence of the phonon linewidth was observed for $E_{2g}$ mode in MgB$_2$, and was explained with the EPC-induced anharmonicity\,\cite{ponosov17,novko20b}.

\begin{figure}[!t]
\centering
\includegraphics[width=0.495\textwidth]{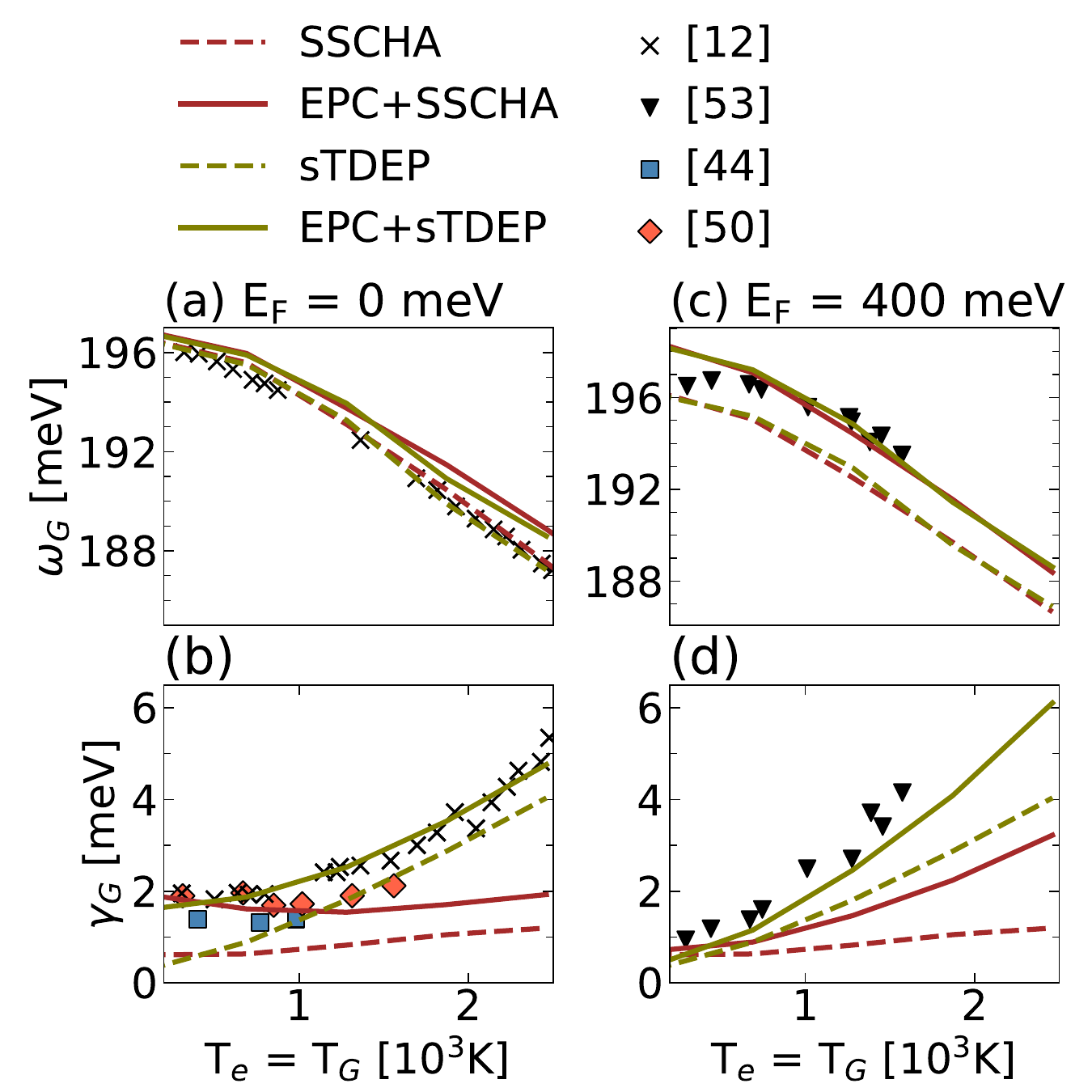}
\caption{\textbf{Total temperature dependence of the G phonon mode in equilibrium graphene.} For the two different doping regimes, indicated at the top of the respective panels with the corresponding Fermi energies, we show the total EPC and anharmonicity-induced G mode (a),(c) frequency renormalization and (b),(d) linewidth as a function of temperature. We compare anharmonic effects obtained with sTDEP and SSCHA. (a),(c) Anharmonic effects have a prevailing temperature dependence at smaller temperatures shown here and determine the G mode frequency decrease with temperature observed in experiments~\cite{MONTAGNAC13,Berciaud2010}.  (b),(d) Linewidth increases with temperature in accordance with the observations in Ref.~\citenum{MONTAGNAC13,liu19,chae10,Berciaud2010}.
} 
\label{fig:fig3b}
\end{figure}

In the remaining two panels of Fig.~\ref{fig:fig3}, we show the doping dependence of the linewidth $\gamma_G$ and inverse lifetime of electron-hole pairs $1/\tau_{\rm op}(\omega_G)$ at 300 K. In panel (g) we show how the intraband contribution to the linewidth becomes increasingly more important with doping in relative terms. This result also explains the decrease and subsequent increase of phonon linewidth $\gamma_G$, as a function of graphene doping with FeCl$_3$\cite{chen23}. The increase as a function of doping was explained in Ref.\,\citenum{chen23} in terms of electronic state broadening due to electron-electron scatterings in Raman process, however, here we show that the EPC-related scatterings could also explain this behavior. In Fig.~\ref{fig:fig3}(h) we show the linear dependence of the electron-hole inverse lifetime on doping. The slope of the curve amounts to 0.027, similar to the estimated value of 0.021 reported in Ref.~\citenum{chen23} for intraband EPC-related transitions near $E_{F}$. 

Having established the important role of the dynamical intraband contribution to the phonon self-energy at higher temperatures and dopings, we simulate the full Raman spectrum of graphene, with both the EPC and standard anharmonic effects obtained with SSCHA and TDEP. Note that here in the main text we report the TDEP anharmonic results obtained with stochastic sampling of the potential energy surface, i.e., sTDEP, while in the Supplementary Figure 2 we show and compare the TDEP results based on MD and PiMD.

Results are presented in Fig.~\ref{fig:fig3b} at a smaller temperature range than in Fig.~\ref{fig:fig3}. The chosen dopings are compared to Raman experiments done on pristine graphene and graphite in panels (a)-(b)~\cite{MONTAGNAC13,liu19,chae10} and doped graphene on SiO$_2$ in panels (c)-(d)~\cite{Berciaud2010}. 
Frequency dependence is largely determined by the anharmonic effects for both doping regimes, since the intraband EPC effects start contributing at larger temperatures. With both sTDEP and SSCHA we obtain a fairly good agreement with the temperature frequency trend observed in experiments for both cases. The SSCHA and sTDEP results greatly overlap, proving the equivalence of the two methods. Note that the full results, which include both EPC effects and standard anharmonicity, slightly overestimate the experimental values for the undoped graphene (by $0.5-1$\,meV), however, it is important to observe that the difference between our results and experiments is within the same scale as the difference between various experimental results. Moreover, the shift of our result with respoct to the experimental values can also be attributed to the fact that our DFPT calculation does not account for lattice thermal expansion. Lattice expansion would lead to a slight reduction of the phonon frequencies.

The importance of the proper inclusion of EPC effects along with the standard anharmonicity is more pronounced for the G mode linewidth. With SSCHA, only three-phonon contributions are included for the linewidths. In agreement with Ref.~\citenum{Han22}, we find that these three-phonon contributions are rather small, and even together with the much larger EPC effects the experimental linewidth values are still underestimated. In Supplementary Figure 3 we show that the three-phonon contributions as obtained with SSCHA and sTDEP are quite similar. With TDEP we are able to account also for the four-phonon contribution that is an order of magnitude larger than the three-phonon term (see Supplementary Figures 3 and 4), leading to an excellent agreement with the experimental linewidths. 
This may raise the question of the importance of even higher orders. However, the larger magnitude of four-phonon compared to three-phonon terms is not due to a higher energy contribution of fourth-order anharmonicity but to the mirror-plane symmetry of graphene which forbids any interaction involving an odd number of flexural phonon. This sum rule restricts drastically the number of available three-phonon processes, leading to its lower contribution, while four-phonon term are less affected. Since forces predicted by the anharmonic potential up to fourth-order allows for an accurate reproduction of the full anharmonic forces, as shown in Supplementary Figure 5, we should expect higher order terms to be mainly negligible.
In Supplementary Figures 4 and 6 we show that the inclusion of the quantum thermal expansion significantly impacts the four-phonon contribution and is important to reach the final agreement with the experiments. A classical approach in the volume expansion calculation leads to an overestimation of the phonon linewidth.

\begin{figure}[!b]
\includegraphics[width=0.5\textwidth]{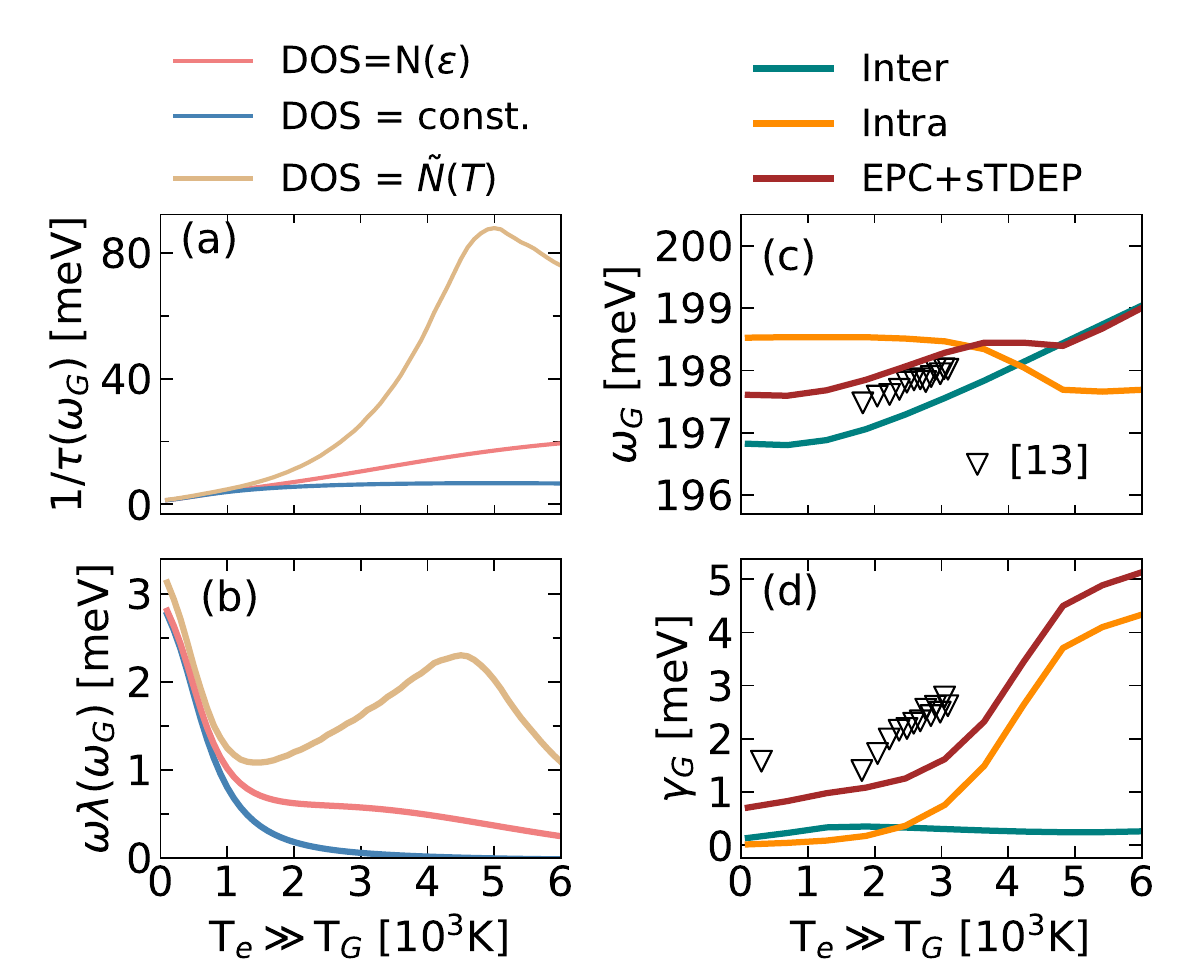}
\caption{\textbf{Phonon dynamics in non-equilibrium graphene.} (a),(b) Electron-hole pair and (c),(d) G mode properties out of equilibrium for $E_{F}=250$\,meV. Phonon temperature is kept constant (300K). The effects of changing the electronic temperature are visible in (a),(b) the electron-hole pair self-energy through the temperature dependent quasiparticle DOS. Both the G mode frequency (c) and its linewidth (d) increase with temperature, as reported in Ref.~\citenum{ferrante18}. 
} 
\label{fig:fig4}
\end{figure}

\textbf{Out-of-equilibrium conditions.} We now turn to a slightly different scenario achieved in ultrafast Raman experiments. The main distinction of the ultrafast regime with respect to the continuous wave measurements is that thermal equilibrium between the electron and phonon subsystems is not achieved\,\cite{yan09,ferrante18}. In one particular case, where $\text{T}_e \gg \text{T}_{G}$, we expect the EPC to play the leading role, as anharmonic effects are not driven by the electronic temperature. We employ the same methodology as for the equilibrium case, but instead of setting the G mode temperature equal to the electronic one, we keep the G mode at 300 K, and vary the electronic temperature T$_e$. In order to simulate the experiment done in Ref.~\citenum{ferrante18}, we perform our calculations on doped graphene with $E_F=250$\,meV. In Fig.~\ref{fig:fig4} we show the inverse lifetime of electron-hole pairs and their energy renormalization, calculated by progressively more complete approaches. As for the equilibrium case, taking into account the temperature-dependent quasiparticle DOS leads to the largest contributions to both quantities. However, for $\text{T}_e \gg \text{T}_G$, an order of magnitude is lost in comparison with respect to results presented in  Fig.~\ref{fig:fig2}, due to the large magnitude of the Bose-Einstein distribution factors for high phonon temperatures. The G mode frequency dependence in the non-equilibrium regime is now almost entirely the result of the interband phonon self-energy. The experimentally observed trend is reversed from the equilibrium one (i.e., increase of the frequency with increasing temperature)\,\cite{yan09,ishioka08,ferrante18}. Our calculations explain that this is due to the decreased impact of the anharmonic and EPC intraband contribution, and reproduce the slope very well. 
The temperature dependence of the linewidth does not change qualitatively in comparison with the equilibrium case, but its magnitude is reduced. The key role belongs to the intraband EPC contribution. We consider EPC to be the driving mechanism behind the temperature dependence in the nonequilibrium case, making the accurate calculation of the EPC phonon self-energy even more essential.

\section*{Discussion}
In this work, we revisit the EPC theory in graphene and reveal its effect on the temperature-dependent properties of the G mode Raman spectral features. The long-wavelength phonons are screened by the first-order interband NA coupling and the higher-order dynamical intraband term, which can be seen as electron-mediated phonon-phonon coupling. The latter effect is essentially anharmonic and, in combination with conventional anharmonicity, captures the full temperature dependence of G mode frequency and linewidth, in line with  experimental results. We believe this type of approach is crucial to describe Raman experiments for materials with significant coupling of electrons to Raman active modes. 
In graphene we find that the EPC-induced and conventional anharmonicities play different roles in different doping regimes. In undoped samples, standard anharmoncity is the dominant contribution to the phonon dynamics, while for the moderately- to highly-doped graphene, the electron-mediated phonon-phonon coupling is crucial. The latter EPC effect is also essential to understand Raman features in non-equilibrium graphene. As for the lattice-driven anharmonicity, by comparing SSCHA and TDEP calculations, we show that four-phonon scattering contributions are more important than the three-phonon scattering for the G phonon linewidth, and that the quantum thermal expansion effects are crucial for obtaining the right values for the four-phonon part.

In conclusion, we emphasize that the present theory could be applied to any material with strong non-adiabatic effects\,\cite{nina23}, such as doped layered materials like transition metal dichalcogenides\,\citenum{sohier19}, hole-doped diamond\,\cite{caruso17}, or MgB$_2$-related compounds\,\cite{yu22}. The present EPC effects might also explain the non-monotonous linewidth behavior in Weyl semimetals, as the standard anharmonic effects are unable to describe the observed temperature dependence\,\cite{Coulter19,wulferding20,osterhoudt21,yu23,cheng24}. In some CDW materials, it could be the leading mechanism behind the CDW transition~\cite{varma83} and it could also be important for strongly coupled or soft phonons away from the Brillouin zone center\,\cite{flicker15}. In that case, it could provide a leading contribution to the formation of the gap or pseudogap\,\cite{yoshiyama86}, as well as to the value of the CDW transition temperature\,\cite{varma83}. Furthermore, our EPC theory could explain the non-monotonous temperature dependence of the phonon linewidth observed for the pristine Ru(0001) surface\,\cite{jardine24}.

\section*{Methods}

\textbf{Electron-phonon contribution.} From first principles we calculate electron and phonon properties with \textsc{QUANTUM ESPRESSO}~\cite{qe, Giannozzi2017, Giannozzi2020} and EPC with the EPW code~\cite{Giustino2007a, Noffsinger2010, Ponce2016}. We use a norm-conserving scalar pseudopotential from \textsc{PseudoDojo}~\cite{VanSetten2018} with an energy cutoff of 100 Ry. The relaxed lattice constant is $2.448\,\text{\AA}$ and the periodic graphene planes are separated by $12\,\text{\AA}$. We use the Fermi-Dirac smearing with $T = 800$\,K.

For the full Brillouin zone phonon calculation, needed for the Eliashberg function calculation, we use a uniform coarse $48 \times 48$ k-mesh.  The phonon calculation is done on a uniform coarse $24\times 24$ q-mesh. 
For the EPC calculation, we use maximally localized Wannier functions~\cite{Marzari2012}, with five initial projections corresponding to one sp$^2$ orbital and two p$_z$ orbitals on the C atom sites. The resulting Wannier functions lie on top of the two C atoms from the unit cell and on the bond centers. Smearing in the EPW calculation is set to 30\,meV, while the electronic temperature in the Fermi-Dirac distribution functions is set to 800 K (this stabilizes the DFT and DFPT calculations as well). The fine k- and q-meshes are $400\times 400$ and $200 \times200$, respectively.

For the $\Gamma$ point phonon and EPC calculations, we sample the k-space more densely using a $96 \times 96$ coarse mesh. The electronic temperature in the Fermi-Dirac distribution functions varies in our calculations. The fine k-mesh for the G phonon is then set to $1000 \times 1000$.

\textbf{Anharmonic contribution.} SSCHA calculations are performed in a $9\times 9$ supercell. The value of Kong-Liu threshold ratio is set to 0.5 and we use up to 6000 random configurations in order to achieve convergence.

The TDEP extractions of interatomic force constants are performed in a $10\times10$ supercell, following MD simulations done with LAMMPS~\cite{Thompson2022} using a Langevin thermostat and a machine-learning interatomic potential (MLIP see below).
For each temperature, four $400$\,ps MD runs in the NVT ensemble were performed and gathered, to ensure a good sampling of the canonical ensemble.
The cutoff for the second order force constants were set to half the simulation cell in the plane. For the third and fourth order interatomic force constant cutoffs we used $7.0$ and $2.6\,\text{\AA}$, respectively.
Phonon lifetimes due to phonon interactions were computed using a $128\times128$ q-points grid when only three phonon interactions are considered and a $48\times48$ q-points grid with fourth order (both are well converged).

\textbf{Construction of a machine-learning interatomic potential.} The MLIP used for molecular dynamics and TDEP was constructed following the framework of the Moment Tensor Potential~\cite{Novikov2021}.
The dataset was built self-consistently, with configurations generated on-the-fly from a molecular dynamics simulations driven by a pre-trained MLIP.
In this MD, a new configuration is extracted every 2\,ps, with energy, forces and stress computed using DFT before being added to the dataset.
At each restart of the MD simulation, the MLIP is updated.
To ensure a good description of the system in the whole temperature range, at each step the temperature of the thermostat in the MD simulation was chosen randomly between 100\,K and 3000\,K.
The simulation cell was allowed to expand anisotropically in the plane using a barostat set to 0 GPa, in order to capture the thermal expansion of the system.

The MD simulations were performed with the LAMMPS package~\cite{Novikov2021} while the first-principles computation of energy, forces and stress used in the MLIP training were done with the Abinit DFT code~\cite{Gonze2020}.



\begin{acknowledgments}
N.G.E., I.L., and D.N. acknowledge financial support from the Croatian Science Foundation (Grant no. UIP-2019-04-6869 and UIP-2020-02-5675) and from the European Regional Development Fund for the ``Center of Excellence for Advanced Materials and Sensing Devices'' (Grant No. KK.01.1.1.01.0001), as well as from the project "Podizanje znanstvene izvrsnosti Centra za napredne laserske tehnike (CALTboost)" financed by the European Union through the National Recovery and Resilience Plan 2021-2026 (NRPP).
J.P.B., A.C. and M.J.V. acknowledge computing time from a PRACE award granting access to Discoverer at SofiaTech in Bulgaria (OptoSpin project id. 2020225411), 
EuroHPC (Extreme grant EHPC-EXT-2023E02-050) on
Marenostrum5 at BSC, Spain, by the CECI (FRS-FNRS Belgium Grant No. 2.5020.11), as well as the Lucia Tier-1 of the F$\text{\'e}$d$\text{\'e}$ration Wallonie-Bruxelles (Walloon).
They also acknowledge financial support from ARC project DREAMS (G.A. 21/25-11) funded by Federation Wallonie Bruxelles and ULiege; the EUSpecLab MSCA DTN network funded by EU Horizon Europe (G.A. 101073486); and the Excellence of Science project CONNECT (G.A. 40007563) funded by FWO and FNRS.
\end{acknowledgments}




\bibliography{ref}

%
%

\end{document}